\documentclass[structabstract]{aa}  
\usepackage{graphicx}
\usepackage{natbib}
\usepackage{color}
\usepackage{url}

\usepackage{amsmath}
\usepackage{txfonts}

\bibpunct{(}{)}{;}{a}{}{,}
\begin{document}

\title{Variations of the solar granulation motions with height using the GOLF/SoHO experiment}

\author{S. Lefebvre
	\inst{1}
	\and R. A. Garc\'ia\inst{1}
	\and S. J. Jim\'enez-Reyes\inst{2}
	\and S. Turck-Chi\`eze\inst{1}
	\and S. Mathur\inst{1, 2}
	}

\institute{Laboratoire AIM, CEA/DSM-CNRS-Universit\'e Paris Diderot; CEA, IRFU, SAp, centre de Saclay, F-91191, Gif-sur-Yvette, France\\
	\email{sandrine.lefebvre@cea.fr; rgarcia@cea.fr}
        \and
	Instituto de Astrof\'isica de Canarias, 38205, La Laguna, Tenerife, Spain\\
        }


\abstract
{Below 1 mHz, the power spectrum of  helioseismic velocity measurements is dominated by the spectrum of convective motions (granulation and supergranulation) making it difficult to detect the low-order acoustic modes and the gravity modes.}
{We want to better understand the behavior of solar granulation as a function of the observing height in the solar atmosphere and with magnetic activity during solar cycle 23. }
{We analyze the Power Spectral Density (PSD) of eleven years of GOLF/SOHO velocity-time series using a Harvey-type model to characterize the properties of the convective motions in the solar oscillation power spectrum. We study then the evolution of the granulation with the altitude in the solar atmosphere and with the solar activity.}
{First, we show that the traditional use of a lorentzian profile to fit the envelope of the $p$ modes is not well suitable for GOLF data. Indeed, to properly model the solar spectrum, we need a second lorentzian profile. Second, we show that the granulation clearly evolves with the height in the photosphere but does not present any significant variation with the activity cycle.}
{}


\maketitle

\section{Introduction}
\label{Intro}

The photosphere, the visible layer of the Sun, is the location where the energy transport previously dominated by convection and turbulence is largely done by radiation with an optical depth of $2/3$. The gas is visible in the form of granules, that penetrate inside the stable photosphere. These granules, as well as other larger structures like the mesogranules or the supergranules, are the manifestation of  the different spatial scales of the convective motions occurring in this region of the Sun \citep{Zahn87,Roudier91,Espagnet93}.

The study of the granulation is particularly important in helioseismology because, on the one hand, it excites the so-called 5-min oscillations, i.e. the acoustic (p) modes, and, on the other hand, it dominates the power spectrum at low frequencies  preventing the detection of low-order p modes. Indeed, in the case of velocity measurements, the lower detection limit of acoustic modes is established around 1 mHz \citep{Garcia01,Garcia04a,Broomhall07}. To progress in the detection of such modes as well as to increase the detection probability of gravity modes  \citep{Appourchaux00,Gabriel02,Turck04,Garcia07,Mathur07} we need to better characterize the properties of the granulation in order to reduce, if possible, their impact on the helioseismic  measurements. The coming GOLF-NG instrument will soon address this problem \citep{Turck06}. This new-generation instrument should improve the signal-to-noise (S/N) ratio of low-frequency modes by measuring the Doppler velocity at different heights in the solar atmosphere. It would thus benefit from the reduction in the coherence of the granulation with the atmospheric altitude \citep{Garcia04b}.

In this paper, we analyze the  Power Spectral Density (PSD) of velocity time sub-series from the GOLF\footnote{Global Oscillation at Low Frequencies \citep{Gabriel95}} instrument on board SOHO\footnote{SOlar and Heliospheric Observatory \citep{Domingo95}} which is a solar disk integrated resonant spectrometer. With such an instrument, we can already study the mean behaviour of the solar granulation in some range of the atmosphere.  So this work  comes in complement to  the study of \citet{Espagnet95} where the authors found that the photosphere is highly structured with two distinct layers below and above about 90 km. With GOLF, and the technique used in \citet{Jimenez07}, we are able to study, without spatial resolution, a region located between 250 up to 550 km above the photosphere.  We show that the granulation evolves with the height in the photosphere and that the granules tend to have shorter lifetimes with a weaker velocity when higher in the atmosphere.

Section \ref{Analysis} is devoted to the data analysis, with first a brief summary of the velocity calibration procedure of the GOLF signal and secondly a description of the fitting procedure of the power spectra in more details. Section \ref{Results} is dedicated to a detailed study of the granulation motions and its evolution with time during the solar cycle. Finally, last section (Sect. \ref{Discussion}) will emphasize the main results of this paper and will anticipate on future incoming works.

\section{Data analysis}
\label{Analysis}

\subsection{Data}
\label{Data}

GOLF is a resonant scattering spectrophotometer that measures the Doppler shift of the neutral sodium doublet ($D_1$ at $\lambda$= 589.6 nm and $D_2$ at $\lambda$ = 589.0 nm). 

\begin{figure}[htbp]
\centering
	\includegraphics[width=9cm]{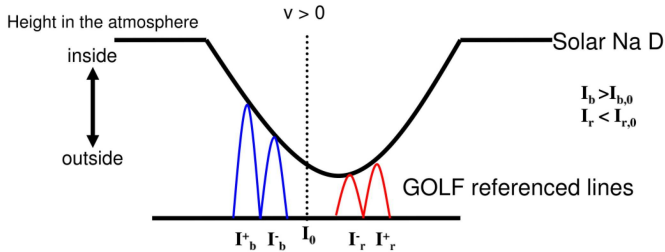}
  	\caption{Representation of the GOLF measurements. The Solar Na D line is displaced from the rest position due to a positive velocity ($V>0$). The BW (blue wing) measures higher in the line which means deeper (close to the photosphere) in the solar atmosphere. The RW (red wing) measures closer to the bottom of the line, i.e. upper in the solar atmosphere.}
\label{fig0}
\end{figure}

In the present study, we use time series of this experiment calibrated into velocity \citep{Garcia05}. In this instrument, the light coming from the solar sodium absorption line (half-width ~500 m\AA) traverses a sodium vapour cell, placed in a longitudinal magnetic field of $\sim$ 5000 Gauss, where it is absorbed and re-emitted in all directions. This scattered light is symmetrically split into its Zeeman components allowing a measurement on either side on the wings of the solar absorption profile. The scattered photons are collected by two photomultiplier tubes. Due to a malfunction of the polarization subsystem that switches between the two wings of the sodium doublet, and to ensure a 100\% of duty cycle, a single-wing working cycle was selected since April 11, 1996. Three different periods can be distinguished 
during the operations of the instrument: from April 1996 to June 25, 1998 GOLF has been observing on the blue wing of the sodium profile; after the SoHO recovery mission in September 1998, GOLF was restarted on the red wing up to November 18, 2002, when it was switched back to the blue-wing configuration. Since then, it has been running, unchanged, in this configuration.
 
The main velocity field ($\pm$0.5 km/s) measured by GOLF is the SoHO orbital velocity. Due to this movement, the working points on the sodium profile change with a 1-year periodicity. Thus, GOLF is measuring at different altitudes in the solar atmosphere (see Figure~\ref{fig0}). Indeed when the orbital velocity $V_{orb}$ increases, the measurements taken on the blue wing are closer to the photosphere, whereas on the red wing the measurements are further from the photosphere \citep{Garcia04b}.
 
The transformation from the observed counting rates to an intensity-like variable suitable for a velocity scaling includes two major steps: 
firstly, corrections to the raw measurements to remove instrumental effects; secondly, compensation for the Sun-spacecraft variable distance.
Then, before calibrating the data into velocity, corrections of some instrumental effects are applied to the counting rates  including those applied to the photomultipliers, such as dead time, ageing of the phototubes and high voltage perturbations, as well as sodium cell stem, photocathodes, and filter temperature corrections (see \citet{Garcia05} for further details). In particular, in the second period of the blue-wing measurements, the correction of the photomultipliers temperature was no longer valid and it has been removed from the calibration pipeline. As a consequence, there is a small non-corrected offset in this last segment of data compared to the first period. Lastly, the calibration introduces a smooth high-pass filter with a cut-off frequency at about $1\times10^{-6}$ Hz that produces an effect till the region of few days.  To check the independence of our results from the calibration procedure, we have repeated the analysis using another velocity calibration procedure developed by \citet{Ulrich00} which involves a different set of corrections. The results were qualitatively the same using both calibration methods.

GOLF has been working for nearly 4200 days, a long period of time that allows a detailed study of the temporal evolution of its power spectrum during the whole solar cycle 23. To understand it and, in particular, the granulation, we are going to use two other quantities:
\begin{itemize}
	\item The MPSI (Magnetic Plage Strength Index), developed by \citet{Chapman86,Ulrich91,Parker98} which can be used as a proxy for the magnetic activity. This index is determined from the 150-foot solar tower magnetograms by measuring the sum (in all pixels) of the absolute value of the magnetic field strengths between 10 and 100 gauss. This number is then divided by the total number of pixels (regardless of magnetic field strength) in the magnetogram (see \url{http://www.astro.ucla.edu/~obs/150_data.html}).
	\item The GOLF observational height in the solar atmosphere computed from line-of-sight velocity taking into account the main 
	displacement of the Na lines from the rest position.  We use the procedure described in \citet{Jimenez07} to translate these velocities into a value of height in the solar atmosphere.
\end{itemize}

It is important to remember that GOLF measures Doppler shifts of the Na line. Thus, it is mostly sensitive to a mixture of the vertical velocity towards the disk center and the horizontal velocity towards the solar limb. In fact, due to the single-wing configuration, the highest sensitivity will not be at the disk center but shifted towards the East or the West depending on the wing where the measurements are done (see  Figure 3 of  \citet{GarRoc1998}).

\subsection{Fitting procedure}
\label{Fitting}

The original GOLF time series are sampled every 20 s and we used series of 91.25 days shifted every 22.81 days. It is important to note that in the statistical analysis we will use independent subseries only, while in the figures we will plot all of them. This length of data has been chosen as a good trade-off between the frequency resolution and the number of independent series needed to correctly track the variations in the height of the atmosphere during a year. Fast Fourier Transform (FFT)  techniques were used to compute each individual spectrum. The normalization used in the PSD is the so-called ``one-sided'' power spectral density \citep{Press92}. A first selection over the spectra was made. We considered only spectra with a duty cycle above 85\%, and we eliminated 3 series in september, october and november 2002 --just before coming back to the blue wing-- because they contain a mix of both wings.

\begin{figure*}[t]
\centering
\begin{tabular}{cc}
	\includegraphics[width=8.2cm]{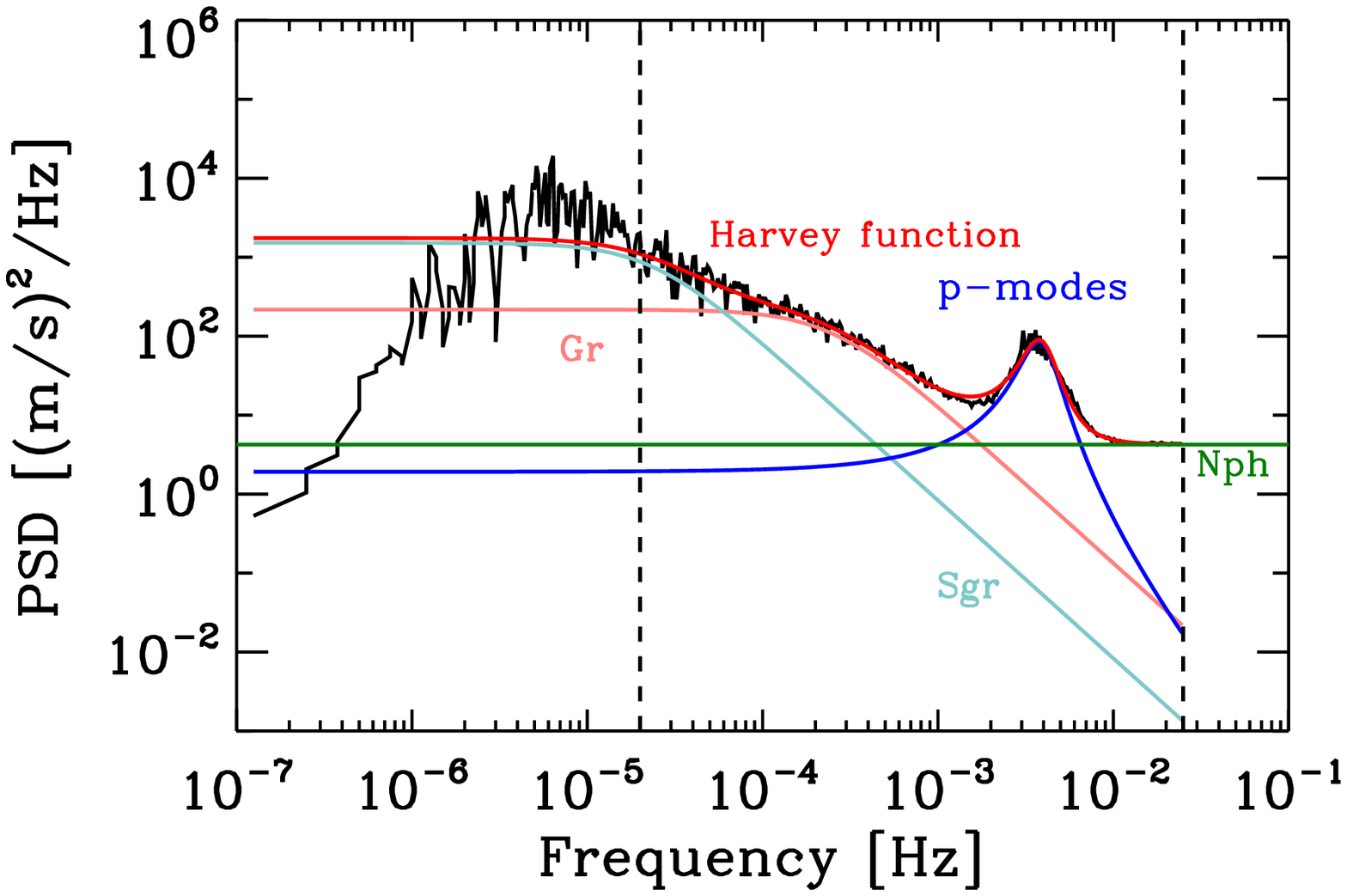} &
  	\includegraphics[width=8.2cm]{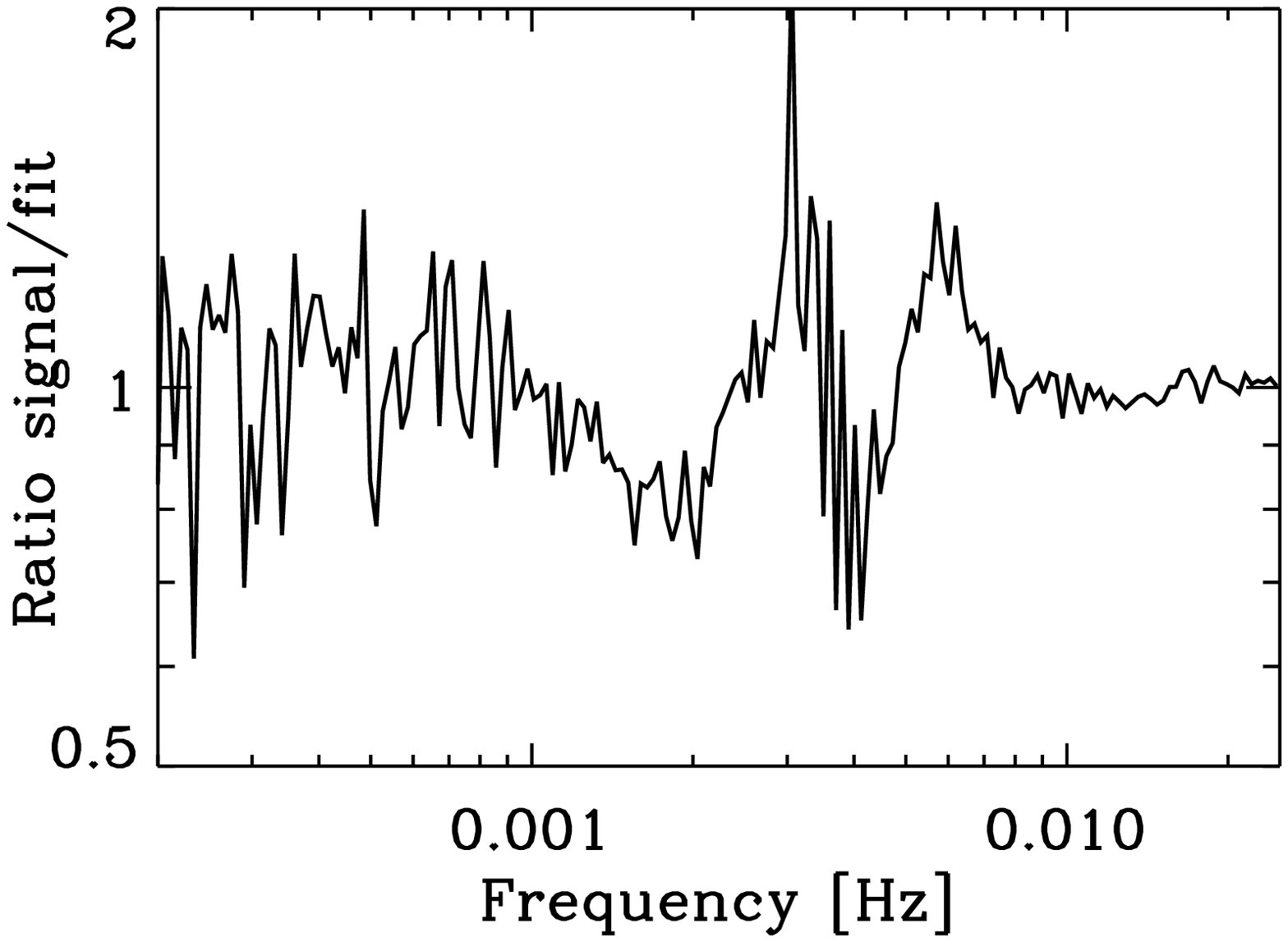} \\
	\includegraphics[width=8.2cm]{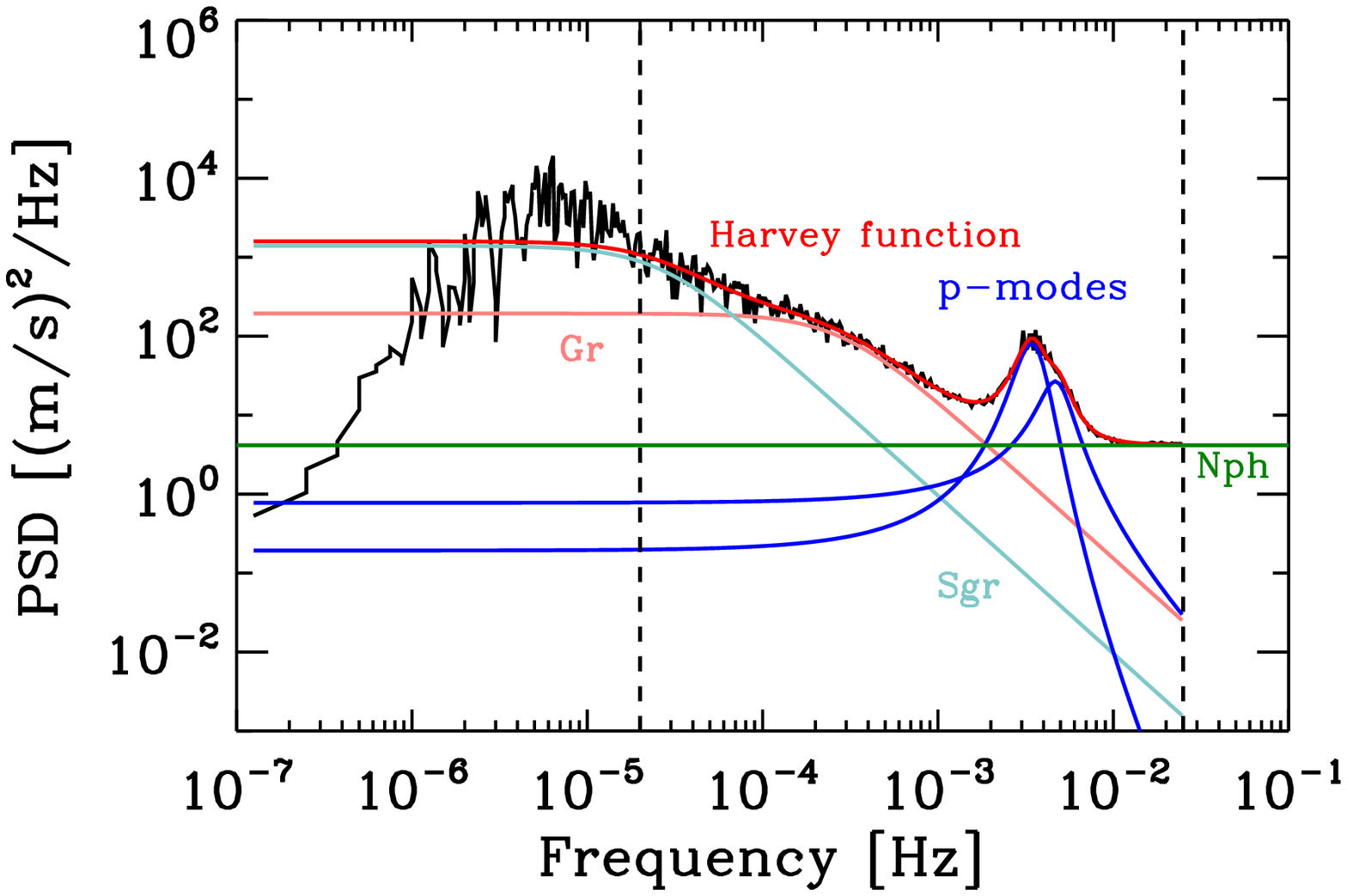} &
  	\includegraphics[width=8.2cm]{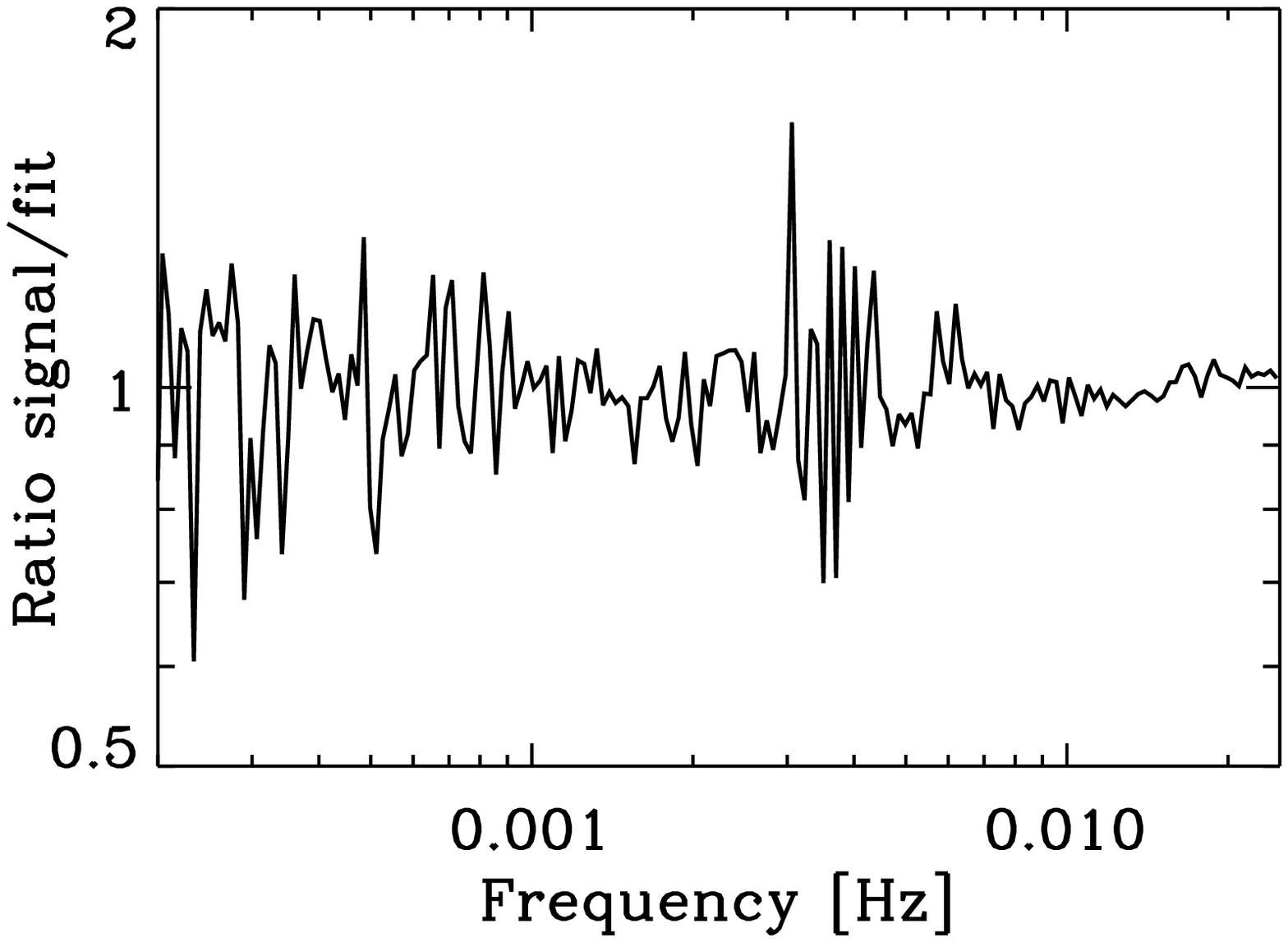} \\
\end{tabular}
  	\caption{Results of two different type of fits, as explained in the text, applied to a GOLF spectrum of an arbitrary taken subseries of 91.25 days long. Top: Left, PSD with a fit using 8 parameters (one lorentzian) to adjust the $p$-mode envelope; Right, ratio between the PSD and the fit around the envelope of $p$-modes. Bottom: Left, PSD with a fit using 11 parameters (two lorentzians) to adjust the $p$-mode envelope; Right, ratio between the PSD and the fit around the envelope of $p$-modes.  The dashed lines represent the limits inside which the fit is performed. The color used for the different fits are: gray for the super granulation contribution, magenta for the granulation contribution, blue for the $p$-mode envelope, green for the noise and red for the harvey function (the sum of the granulation and supergranulation contributions).}
  	\label{fig1}
\end{figure*}

The velocity power spectral density (PSD) can be explained by a model in which each source of solar convective motions is described by an empirical law initially proposed by Harvey \citep{Harvey85,Harvey93} and corresponding to an exponentially decaying time function, plus one or two Lorentzian functions for the envelope of $p$-modes \citep{Vasquez02} and a constant for the photon noise. The fit of the spectra, including both non-periodic and periodic components, are thus expressed by:

\begin{equation}
\begin{split}
P(\nu) 
& = N_{ph} + \sum^{N}_{i=1}\frac{4\sigma_i^2\tau_i}{1+(2\pi\nu\tau_i)^{b_i}} \\ 
& + \sum^{M}_{j=1}A_j\left[\frac{\Gamma_j^2}{(\nu-\nu_{0_j})^2+\Gamma_j^2}\right]^{c_j} \\
\end{split}
\end{equation}
where
\begin{itemize}
	\item $P(\nu)$ is the power spectral density; 
	\item $N_{ph}$ is the photon noise in units of $(m/s)^2/Hz$;
	\item $i$ corresponds to the non-periodic velocity fields;
	\item $j$ corresponds to the periodic components;
	\item $\sigma_i$ and $\tau_i$ are respectively the rms-velocity ($m/s$) and the characteristic time ($s$) of the $i$-th background component (the limit of the first sum $N$ varies depending on the number of non-periodic background components of the spectrum to fit);
	\item $A_j$ and $\nu_{0_j}$ are the power ($(m/s)^2/Hz$) and the central frequency ($Hz$) of the Lorentzian profiles to fit to the periodic components at the higher frequency region of the spectrum, while $\Gamma_j$ sets its width ($Hz$). These $M$ possible peaks to fit can be identified as the so called photospheric or/and the chromospheric component;
	\item finally, $c_j$ (as well as $b_i$) are decay rates.
\end{itemize}

Before fitting the spectra, due to a large number of bins and because the points are compressed in a logarithmic scale at high frequencies, different local averages have been made until the data are equally spaced in the logarithmic scale. This allows to fit the model all along the frequency axis. The fitting is performed in logarithm of the PSD employing a standard non-linear least-square method (Levensberg-Marquardt) \citep{Press92}.

Two non periodic components are fitted for granulation and supergranulation with $b_i=2$. To fit correctly the $p$-mode envelope, it is necessary to use 2 Lorentzian profiles. In preliminary fits, $\Gamma_j$ was obtained close to $1\times10^{-3}$, then we fixed this coefficient to this value for the rest of the fits. In previous work \citep{Regulo02}, an additional factor $(\nu/\nu_{0_j})^{a_j}$ can be found in the Lorentzian term, but  following these authors, we have decided to fix the coefficient $a_j$ to zero in our fits.

We have tried to isolate in the background spectrum different effects that are merged in the data: the dependence with solar activity and the dependence with the observing point in both wings (solar atmospheric depth). Which effect is dominant? In the following, we note BW1 the first period in the blue wing, RW the following red-wing configuration and BW2 the second period in the blue-wing mode. Thus, the fits have 11 coefficients, i.e. 2 coefficients for the granulation as well as for the super granulation, 3 coefficients for each of both Lorentzians used to fit the $p$-mode envelope (the $\Gamma$ factor being fixed), plus the photon noise. If we fit with 8 coefficients, only one Lorentzian is used for the acoustic-mode envelope.

Figure \ref{fig1} shows an example of spectrum where each component is represented by a different curve and the sum of these curves is the total fit (left-hand panels). The ratio between the PSD and the fit is also shown (right-hand panels). We see the difference between a fit with one Lorentzian (8 parameters, top panels) and a fit with two Lorentzians (11 parameters, bottom panels) for the adjustement of the $p$-mode envelope. These plots indicate the necessity of both Lorentzians to correctly adjust the $p$-mode envelope and thus decrease both excess power present near 0.003 and 0.005 Hz in the 8-parameter fit. In the following section, we will focus on the deep analysis of the behavior of the granulation noise. To do so, we model the PSD using only a fit with two Lorentzian profiles (the conclusions obtained for this convective pattern would be qualitatively the same when using a fit with a single Lorentzian). The analysis of the other parameters will be left for future incoming works. However, we note that the fit of the supergranulation is less robust than the fit of the granulation. The supergranulation region is certainly affected by the calibration of the data that filters this region slightly \citep{Garcia05}, and by the small number of points in this region due to a small frequency resolution. If we used non-filtered series of 365 days, the fit of the supergranulation is more stable.

\begin{figure*}[htbp]
	\centering
	\includegraphics[width=17cm]{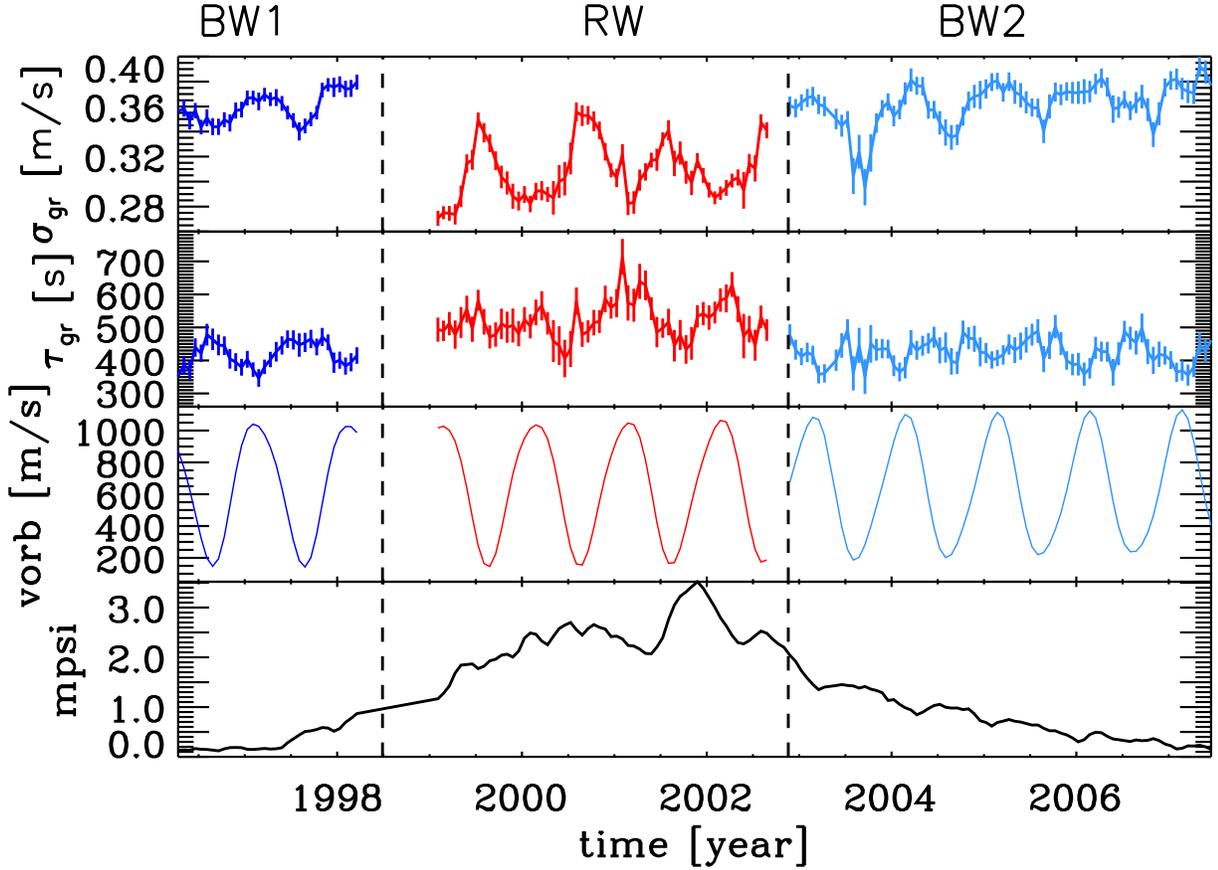} 
  	\caption{Temporal evolution of the granulation ($\sigma$ and $\tau$, top and second panel respectively) of the GOLF PSD during the solar cycle represented by the MPSI index plotted in the bottom panel. The third panel represent the orbital velocity with time. Here, the vertical dotted lines indicate the date of the changes in the GOLF observing configuration.}
  	\label{fig2}
\end{figure*}

\section{Results}
\label{Results}

\subsection{Evolution of the granulation with time}
\label{Time}

Figure \ref{fig2} shows  the temporal evolution of the fitting parameters for the granulation as well as the activity during the last solar cycle represented by the MPSI index.
The rms velocity in each wing is different, with a lower value in the red wing. Besides, the velocity is higher in BW2 than BW1 as a consequence of the change in the calibration of the second blue-wing period when the photocathode temperature correction were removed. Moreover we clearly see the signal dependence with orbital velocity as a one-year modulation, enabling to probe different heights in the atmosphere as we will see in the next section. This one-year modulation has an opposite phase between the blue and red wings, that can be clearly seen on the top panels. The blue wing is directly correlated with the orbital velocity (third pannel) while the red wing is anticorrelated. Moreover, the amplitude is larger in the red wing than in the blue one.

\begin{figure*}[htbp]
	\centering
	\includegraphics[angle=90,width=17cm]{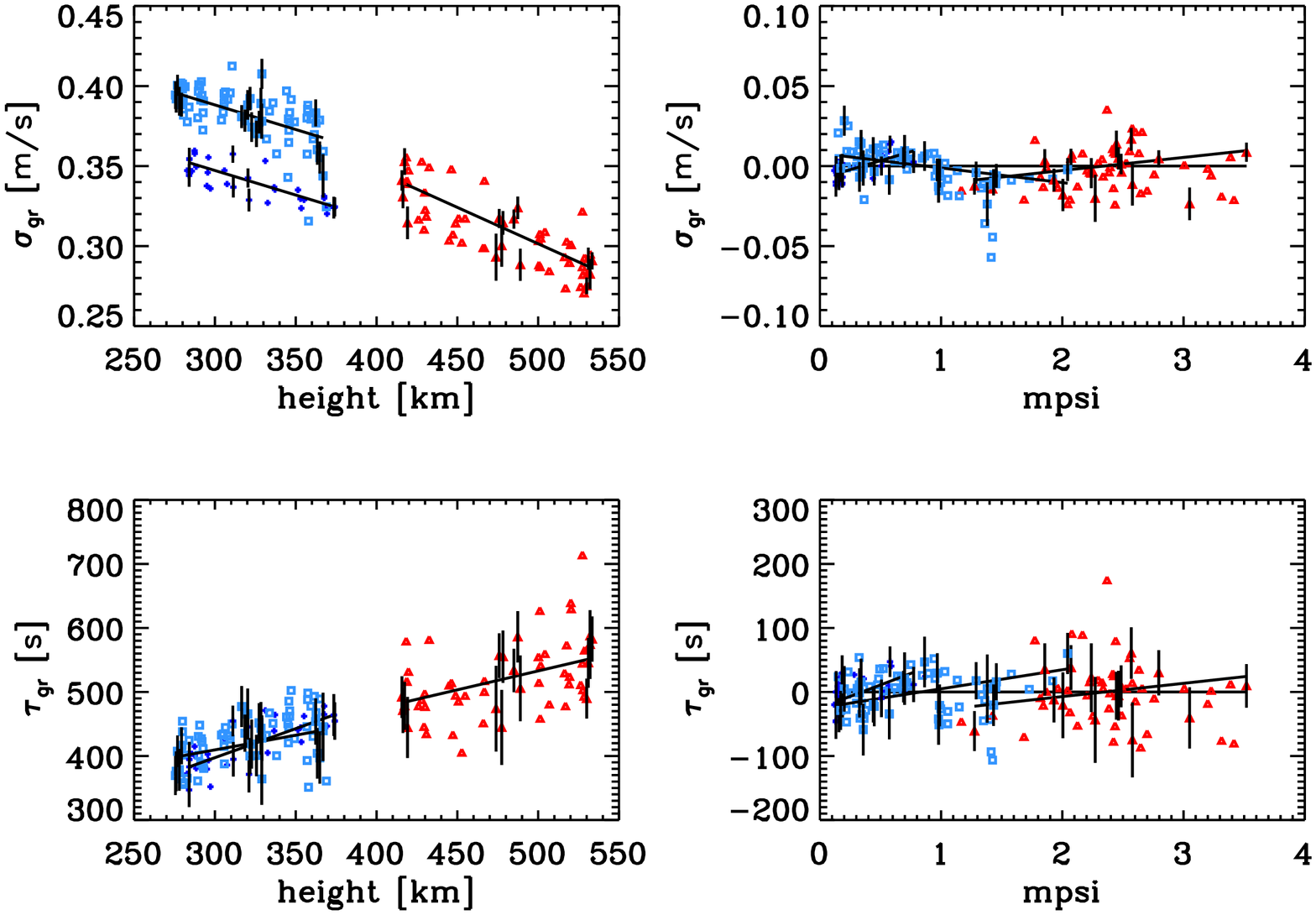} 
  	\caption{Evolution of the granulation velocity and characteristic time versus the height (left) and the MPSI index (right). The linear trend with MPSI has been computed after removing the trend with height. \textcolor{blue}{$+$} and \textcolor{blue}{$\square$}: respectively BW1 and BW2; \textcolor{red}{$\triangle$}: RW. The solid lines in each panel are the best fitting gradient of linear regressions for independent series and for the separate blue- and red-wing data sets. Note the jump between red and blue operating mode. All panels show overlapping data for illustrative purposes, with errors plotted only on the independant data.}
  	\label{fig3}
\end{figure*}

\subsection{Evolution of the granulation with the height and the solar activity}
\label{Height}

The height in the photosphere is computed following the procedure described in detail in \citet{Jimenez07}, and briefly summarized here:
\begin{itemize}
	\item The so-called response functions (hereafter RFs; see \citet{Mein71,Beckers75}) of the Na D1 and D2 Fraunhofer lines are computed: these functions measure the reaction of the line profile when the atmosphere is perturbed locally at a given height. The quiet Sun model C \citep{Vernazza81} was employed as a reference because it produces spectral line profiles (both photospheric and chromospheric) that are generally in good agreement with observations made at low spatial resolution.
	\item For each line we computed the RFs to changes in velocity by applying small perturbations on the model at each optical depth.
	\item Then, the RFs for the full line profiles were convolved with the wavelength filter response of the instrument.
	\item The calculations took into account changes to the line-of-sight velocity at different times of year.
\end{itemize}

Figure \ref{fig3} shows the evolution of the granulation parameters the altitude in the solar atmosphere and the MPSI index. In order to provide a quantitative assessment of the observed changes for the velocity and the characteristic time, we performed a linear regression with each of the parameters: the characteristic of each linear trend is summarized in Table \ref{table1} for the variation with height and in Table \ref{table2} for the variation with the MPSI index. We used data from independent series that did not overlap in time. This was done for each of the blue- and red-wing periods following the same method than in \citet{Jimenez07}. For the evolution with the MPSI index, the trend has been calculated after removing the trend with height. Errors issued from the general fit have been used to ponderate the linear fitting. For the clarity of the figures, only error bars for the independent series have been plotted. The best fit for each of the red and blue data sets are indicated by the black lines in the panels of Figure \ref{fig3}. We also computed the Pearson and Spearman-rank correlation coefficients for the data that are also presented in both tables. In Table \ref{table1}, there are also coefficients for the linear regression with all wings: in the case of the velocity and in order to remove the offset, we have applied an adhoc correction simply by adding (respectively removing) the half of the difference in the velocity intercepts (i.e. $(0.48-0.44)/2$) to BW1 (respectively BW2).

\begin{table*}
\begin{minipage}[t]{\linewidth}
\renewcommand{\footnoterule}{}  
\caption{Linear regression of the granulation with height.}
\label{table1}
\centering
\begin{tabular}{c c c c c c c c c c}
\hline\hline
Epoch	 	& Parameter		& Intercept 		& Gradient 				& Significance 		& $\chi^2$ & N\footnote{Actual numbers of independent data points employed in each of the regression analysis}	& $r_P$\footnote{Pearson correlation coefficient}	& $r_S$\footnote{Spearman rank correlation coefficient}	& $P_S$\footnote{Two-sided probability of Spearman coefficient}\\
		&			&			&					& Of fitted gradient 	&	   &    & & &	\\
\hline
BW1	 	& $\tau_{gr}~(s)$ 	& 125 $\pm$ 102		& 0.91 $\pm$ 0.32			& 2.8$\sigma$	& 1.66	& 8 & 0.71 & 0.55	& 1.60$\times10^{-1}$	\\
		& $\sigma_{gr}~(m/s)$	& 0.44 $\pm$ 0.02	& (-3.06 $\pm$ 0.70)$\times10^{-4}$	& 4.4$\sigma$	& 1.99	& 8 & 0.79 & 0.85	& 6.53$\times10^{-3}$	\\
\hline
RW	 	& $\tau_{gr}~(s)$ 	& 238 $\pm$ 109		& 0.59 $\pm$ 0.23			& 2.6$\sigma$	& 0.94	& 15 & 0.62 & 0.64	& 1.03$\times10^{-2}$	\\
		& $\sigma_{gr}~(m/s)$	& 0.53 $\pm$ 0.02	& (-4.54 $\pm$ 0.43)$\times10^{-4}$ 	& 10.6$\sigma$	& 2.43	& 15 & 0.83 & 0.79	& 4.22$\times10^{-4}$	\\
\hline
BW2	 	& $\tau_{gr}~(s)$ 	& 272 $\pm$ 82		& 0.46 $\pm$ 0.26			& 1.8$\sigma$	& 0.81	& 19 & 0.39 & 0.25	& 3.04$\times10^{-1}$	\\
		& $\sigma_{gr}~(m/s)$	& 0.48 $\pm$ 0.02	& (-3.07 $\pm$ 0.66)$\times10^{-4}$	& 4.7$\sigma$	& 1.50	& 19 & 0.70 & 0.63	& 4.12$\times10^{-3}$	\\
\hline
All wings	& $\tau_{gr}~(s)$ 	& 214 $\pm$ 24		& 0.64 $\pm$ 0.07			& 9.1$\sigma$	& 0.93	& 42 & 0.84 & 0.78	& 1.49$\times10^{-9}$	\\
		& $\sigma_{gr}~(m/s)$	& 0.47 $\pm$ 0.005	& (-3.34 $\pm$ 0.13)$\times10^{-4}$	& 25.7$\sigma$	& 1.95	& 42 & 0.93 & 0.88	& 4.59$\times10^{-15}$	\\
\hline
\end{tabular}
\end{minipage}
\end{table*}

\begin{table*}
\begin{minipage}[t]{\linewidth}
\renewcommand{\footnoterule}{}  
\caption{Linear regression of the granulation (residuals after removing the trend with height) with MPSI index.}
\label{table2}
\centering
\begin{tabular}{c c c c c c c c c c}
\hline\hline
Epoch	 	& Parameter		& Intercept 		& Gradient 				& Significance 		& $\chi^2$ & N\footnote{Actual numbers of independent data points employed in each of the regression analysis}	& $r_P$\footnote{Pearson correlation coefficient}	& $r_S$\footnote{Spearman rank correlation coefficient}	& $P_S$\footnote{Two-sided probability of Spearman coefficient}\\
		&			&			&					& Of fitted gradient 	&	   & & & &	\\
\hline
BW1	 	& $\tau~(s)$ 		& -25 $\pm$ 17				& 72 $\pm$ 41				& 1.8$\sigma$	& 1.15	& 8 & 0.22	& 0.19	& 6.51$\times10^{-1}$	\\
		& $\sigma~(m/s)$	& (-7.88 $\pm$ 4.09)$\times10^{-3}$	& (2.25 $\pm$ 0.97)$\times10^{-2}$	& 2.3$\sigma$	& 1.11	& 8 & 0.56	& 0.60	& 1.20$\times10^{-1}$	\\
\hline
RW	 	& $\tau~(s)$ 		& -49 $\pm$ 43				& 21 $\pm$ 18				& 1.2$\sigma$	& 0.83	& 15 & 0.04	& -0.02	& 9.50$\times10^{-1}$	\\
		& $\sigma~(m/s)$	& (-19.14 $\pm$ 7.39)$\times10^{-3}$	& (0.82 $\pm$ 0.31)$\times10^{-2}$	& 2.6$\sigma$	& 1.88	& 15 & 0.28	& 0.24	& 3.98$\times10^{-1}$	\\
\hline	
BW2	 	& $\tau~(s)$ 		& -25 $\pm$ 15				& 20 $\pm$ 15				& 1.3$\sigma$	& 0.58	& 19 & 0.45	& 0.32	& 1.88$\times10^{-1}$	\\
		& $\sigma~(m/s)$	& (7.96 $\pm$ 3.76)$\times10^{-3}$	& (-9.16 $\pm$ 3.71)$\times10^{-3}$	& 2.5$\sigma$	& 1.14	& 19 & 0.38	& 0.41	& 8.37$\times10^{-2}$ 	\\
\hline
\end{tabular}
\end{minipage}
\end{table*}

The dependence with the depth in the solar atmosphere is well visible. As it was explained above, when the observations are done in the red wing, we are observing higher in the solar atmosphere than when the blue wing is used.
We notice several interesting behaviours:
\begin{itemize}
 	\item For each wing, there is a clear dependence with the height: a decrease of $\sigma_{gr}$ and an increase of $\tau_{gr}$ with the altitude in the photosphere, that can be interpreted as a longer lifetime for granules higher in the atmosphere. There is a strong correlation with height, more important for $\sigma_{gr}$ than for $\tau_{gr}$ (fitted gradients significant at 4.4, 10.6, 4.7$\sigma$ and 2.8, 2.6, 1.8$\sigma$ for each mode operation). For $\sigma_{gr}$, the difference in the intercept between each wing is significant, whereas the error bars for the intercepts for $\tau_{gr}$ are really too big to be significant. It is worth noting that the significance of the linear fit for $\tau_{gr}$ is more important when considering all confused wings.
	\item Concerning the velocity, there is an offset between the two blue wings but the same slope and the same linear trend (slope and offset) for the characteristic time; this is normal because a difference in the temperature calibration between the two wings can affect $\sigma_{gr}$ but not the value of $\tau_{gr}$. The offset in velocity is significant and certainly due to the difference in calibration.
	\item The difference of slope between the blue wing and the red wing for the velocity is weak. It seems that there is a slight break in the profil of $\sigma_{gr}$ with height, near 400 km.
	\item However, there is nothing significant with the MPSI index: the scatter of the residuals with the activity is too big.
\end{itemize}

\section{Conclusion and perspectives}
\label{Discussion}

In this paper we have investigated the vertical structure and time evolution of the solar granulation by means of a novel methodology based on the analysis of the full-disk Sun-as-a-star Doppler velocity observations. Thus we have been able to study the vertical velocity fluctuations and lifetimes of the solar granulation. We have shown that the GOLF PSD can be correctly characterized by our model, a Harvey function with two Lorentzian profiles. 

This work extends the study of \citet{Espagnet95} where they found that the photosphere is highly structured with two distinct layers below and above about 90 km. With GOLF, we  study the photosphere above $\approx$ 280 km and we showed that granules tend to live longer with a weaker velocity when higher in the atmosphere.   Following the results of \citet{Title89}, there is a strong correlation between the granule sizes and lifetime. Therefore, we can conclude that larger granules reach the top of the photosphere, while penetration of small granules decreases with the size. To be more precise, as Figure \ref{fig3} shows a lifetime of about 400-550 s for granules between 250 and 550 km, we can estimate from the article of \citet{Title89} and their figure 21 that the granules in these altitudes have a lifetime-average size of about 1.2 arcsec and a maximum size of about 1.4-1.5 arcsec.

We have found that the granulation rms velocities ($\sigma_{gr}$) are between 0.28 and 0.4 $ms^{-1}$. These values are very different from those obtained from high-resolution measurements which are about 1 $kms^{-1}$.
The difference in spatial resolution is the most likely cause of this difference of three orders of magnitude. 
 
However we did not put in evidence a change with the solar activity. This is consistent with the result of \citet{Jimenez03} that the solar cycle effects are very small compared to the change in the observing height in the photosphere due to the orbital motion. It could be due to the fact that GOLF observes in the more stable part of the sodium lines. However if we do not find any significant variation with the cycle, an other study \citep{Muller07} found a cyclic variation of the contrast of the granules, nearly in phase with the solar cycle, the contrast being smaller at the periods of solar maximum, but no corresponding variation in the scale. 

This work opens two perspectives:  (1) to better understand the evolution of the whole solar convective background during the activity cycle, and specially the evolution of the acoustic-mode envelope presently characterized by the presence of two Lorentzians. We are likely to think that the excess power characterized by the second Lorentzian corresponds to the presence of chromospheric modes, (2) to extend this study to our new instrument GOLF-NG observation. It will cover a larger part of the solar atmosphere thanks to the 8 extractions of the Doppler velocity with a proper determination of their location due to the measurement of both wings and the use of only the D1 line  \citep{Jimenez07, Turck06, Turck08}.

\begin{acknowledgements}

The authors thank all their colleagues (scientists, engineers and technicians) involved with the GOLF instrument aboard SoHO which is a space mission of international cooperation between ESA and NASA. S. Lefebvre is supported by a CNES/GOLF research engineer contract in SAp. S. J. Jim\'enez-Reyes acknowledge partial financial support from Spanish grants PNAyA2007-62651 and the support of European Helio- and Asteroseismology Network (HELAS), a major international collaboration funded by the European Commission's Sixt Framework Programme.

\end{acknowledgements}

\bibliographystyle{aa}
\bibliography{./biblio}

\end{document}